# Collisionless plasma interpenetration in a strong magnetic field for laboratory astrophysics experiments

Ph. Korneev,[1,2,a)] E. d'Humières[1], and V. Tikhonchuk[1]
[1)] *University Bordeaux 1 - CNRS - CEA, CELIA (Centre Lasers Intenses et Applications) UMR 5107, 33400 Talence, France*
[2)] *National Research Nuclear University "MEPhI", 115409, Moscow, Russian Federation*

A theoretical analysis for astrophysics-oriented laser-matter interaction experiments in the presence of a strong ambient magnetic field is presented. It is shown that the plasma collision in the ambient magnetic field implies significant perturbations in the electron density and magnetic field distribution. This transient stage is difficult to observe in astrophysical phenomena, but it could be investigated in laboratory experiments. Analytic models are presented, which are supported by particls-in-cell simulations.

## I. INTRODUCTION

Astrophysical phenomena are of great interest since they proceed in a very large scale and imply the greatest known energies. It is however difficult to understand all the corresponding physics due to extremely large time scales and distances and the impossibility of controlling parameters of interaction. Laboratory experiments and numerical modelling for the astrophysics-related parameters, see, i.e.[1–6], may provide informations complementary to the classical observational astronomy, but such experiments need a substantial deposition of energy in plasma flows and external magnetic fields that become available in high power laser systems[7,8].

Recently, a possibility to create very strong pulsed magnetic fields upto tenths Teslas has been demonstrated in laboratory conditions[9]. They may be used, for example, to model astrophysical magnetized collisionless shocks. The main parameters of such a problem are the magnetization and plasmas velocities[2,3,10]. Rescaling to astrophysical values shows, the subrelativistic plasma flows ($\sim 0.1..0.3$ of the light velocity) and magnetization achievable for these flows with modern magnetic pulsers (about $10^{-2}..10^{-1}$) allow to meet conditions of the Super Novae Remnants (SNR) expanding into a magnetized interstellar medium. The corresponding experiments are upcoming.

Because of a relatively short life time, the laboratory plasmas cannot reproduce long-living structures observed in the Universe. In contrast, it makes possible to study the formation of the collisionless magnetized shocks in great details. Theoretical studies of collisionless shocks show the importance of the magnetic fields at the ion characteristic times, and the particle acceleration in a shock front[4,5,11–16], but one of the most puzzling questions is how the ions are slowed down and how the equilibration in energy between electrons, ions, and magnetic fields occurs.

In this paper a theoretical study of the transient interaction stage in astrophysical-type experiments is presented. We consider colliding plasmas in an ambient magnetic field, corresponding to a generic geometry of an experimental setup. In our study the plasma flows themselves may be magnetized or not, although because of plasmas expansion and magnetic field compression it may happen, that the magnetic fields inside and outside plasmas affect strongly the plasma dynamics. We study the initial stage of interaction when the ions are interpenetrated but not yet slowed down, while the interaction of electrons with the magnetic field creates a front structure that may be a site for the shock formation.

Currently the experiments with electron-ion plasmas can be conducted on high power laser facilities[17], but soon electron-positron pair plasmas may be also available for astrophysical-type experiments[18]. That is why our analysis is essentially related to electron-ion plasmas, but the case of electron-positron plasmas will also be considered.

The structure of the paper is as following: in Section II, we describe a simple hydrodynamic model for cold plasma flows moving towards each other in the presence of the ambient magnetic field. Then, in Section III, the results of PIC simulations are analyzed with the help of proposed hydrodynamical models. Finally we conclude with the discussion of our findings. The cumbersome calculations are detailed in Appendix.

## II. GENERAL EQUATIONS AND SIMPLE HYDRODYNAMIC MODELS

The analytical model considers a cold collisionless plasma described with the hydrodynamic approximation. As we are interested here in the transient stage, the ions are considered as a stationary uniformly charged background. The general set of equations for electrons is

[a)]Electronic mail: korneev@theor.mephi.ru

written below.

## A. Hydrodynamic equations for plasma flows

The electron dynamics is described by the continuity equation related the density $n$ and velocity $\vec{v}$ of each flow:

$$\frac{\partial n}{\partial t} + \operatorname{div} n\vec{v} = 0, \quad (1)$$

and the Euler equation:

$$\frac{\partial \vec{p}}{\partial t} + (\vec{v}\nabla)\vec{p} = -e\vec{E} - e\left[\frac{\vec{v}}{c} \times \vec{B}\right], \quad (2)$$

where $-e$ is the electron charge,

$$\vec{p} = \gamma m_e \vec{v} \quad (3)$$

is electron momentum, $\gamma = 1/\sqrt{1-v^2/c^2}$, $m_e$ is an electron mass, $c$ is the velocity of light, and $\vec{E}$ and $\vec{B}$ are electric and magnetic fields. Then, the set of Maxwell's equations

$$\operatorname{rot}\vec{B} = \frac{4\pi}{c}\vec{j} + \frac{1}{c}\frac{\partial \vec{E}}{\partial t}, \quad (4a)$$

$$\frac{1}{c}\frac{\partial \vec{B}}{\partial t} = -\operatorname{rot}\vec{E}, \quad (4b)$$

$$\operatorname{div}\vec{E} = 4\pi\rho, \quad (4c)$$

$$\operatorname{div}\vec{B} = 0, \quad (4d)$$

is related to hydrodynamics by the current and the charge density that are calculated in a standard way.

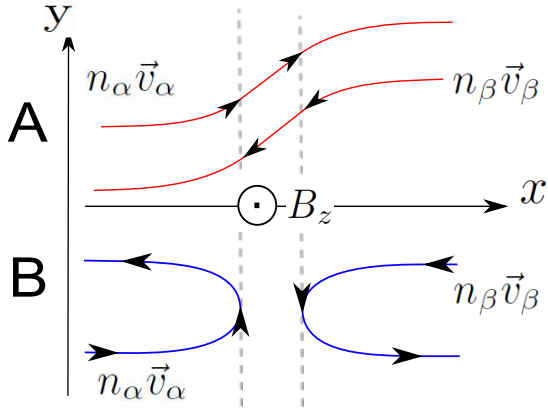

FIG. 1. The scheme for the stationary solutions for electron fluxes in the "compressed magnetic mirror model".

For the qualitative description of main phenomena during the interaction it is useful to find stationary solutions for plasma flows. They may be used as a starting point for the stability analysis or the ion dynamics studies.

We consider stationary flows in the presence of unidirectional (along the $z$-axis) magnetic field, which is normal to the plane $(x,y)$ of flows propagation. This situation corresponds to an experiment of two plasma flows colliding in an ambient magnetic field. As the simplest model, one can imagine two electron-ion flows, propagating from one infinity to another, through the magnetic field barrier ($B_z(\pm\infty) \to B_0$). Two possibilities are shown in figure 1. In the case A the electrons are deflected in the field $\vec{B}$, but penetrate through it, and finally get the same velocity as in the beginning. In contrast, in the case B, the field is strong enough to reflect electrons back with the same velocity in an absolute value, but of the opposite sign. The symmetry of the problem allows both two-flux solutions.

However, a simple reasoning may be found against the possibility of realization of the case A. Indeed, taking into account the symmetry of the problem, one finds that the total electric current in the case A is zero everywhere, which means, according to (4a), that there can not be a magnetic field, which deflects an electron flow in the opposite directions. Note, that this conclusion does not concern more complicated solutions, such as less symmetric, multiflows, or kinetic ones.

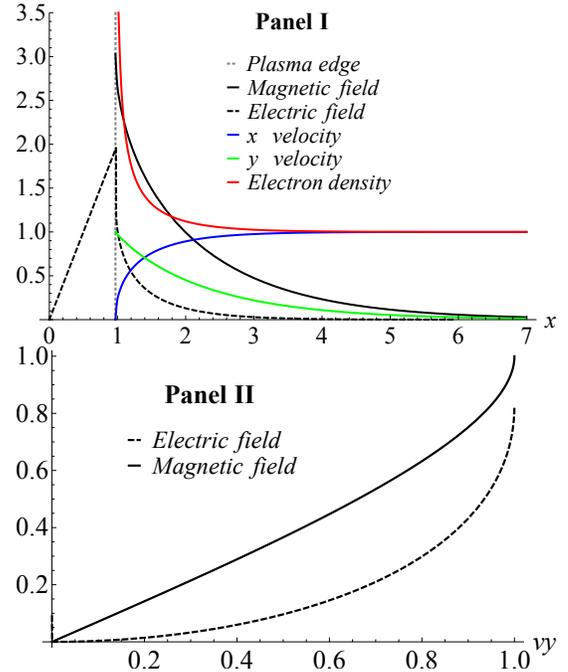

FIG. 2. The approximate solution in the "reflecting mirror" scenario (panel I) and normalized fields $E_x(v_y)/\sqrt{\gamma_0}$ and $B(v_y)/\sqrt{\gamma_0 v_0^2}$ as functions of $v_y$ (panel II) in the case $v_0 = 0.9$, $B_0 = 0$. On the panel I the black solid line is the magnetic field $B(x)$, the black dashed is the electric field $E_x(x)$, the red line is electron density $n(x)$, the blue line is the x-component of the electron velocity $v_x(x)$ and the green line is the y-component of electron velocity $v_y(x)$.

In constrast, the case B in figure 1 corresponds to a stationary situation, where all time dependences in (1), (2), (4) disappear. The symmetry of the problem allows to choose the magnetic field direction along the $z$ axis, to make all the quantities depending only on the $x$ coordinate; then the potential electric field has only the $x$ component, and the current with the velocity have both $x$ and $y$ components. The two oppositely directed flows $\alpha$ and $\beta$ are defined so that $n_\alpha$ and $n_\beta$ are electron densities, $\vec{v}_\alpha$ and $\vec{v}_\beta$ are electron velocities, $n_i$ is the constant ion density, ions are of charges $+e$ for simplicity, and thus in infinity both electron and ion densities are $n_{\alpha,\beta} = n_i \equiv n_0$. At the point $v_x = 0$ the continuity of the flows requires $n_\alpha = n_\beta$. The stationary hydrodynamic equations read

$$\frac{\partial}{\partial x} n_{\alpha,\beta} v_{x;\alpha,\beta} = 0, \tag{5a}$$

$$v_{x;\alpha,\beta} \frac{\partial}{\partial x} \gamma \vec{v}_{\alpha,\beta} = -\frac{e}{m_e}\left(\vec{E} + \left[\frac{\vec{v}_{\alpha,\beta}}{c} \times \vec{B}\right]\right). \tag{5b}$$

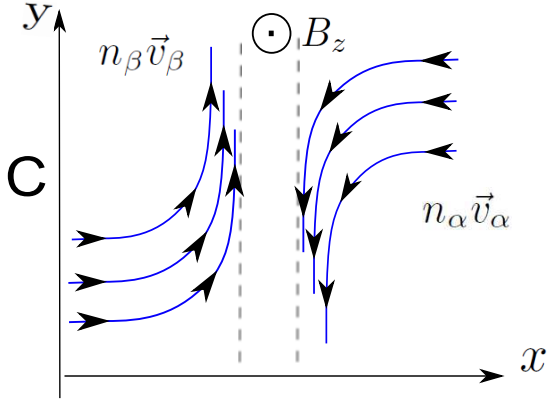

FIG. 3. Quasistationary solution for electron fluxes in the model of the "broken mirror".

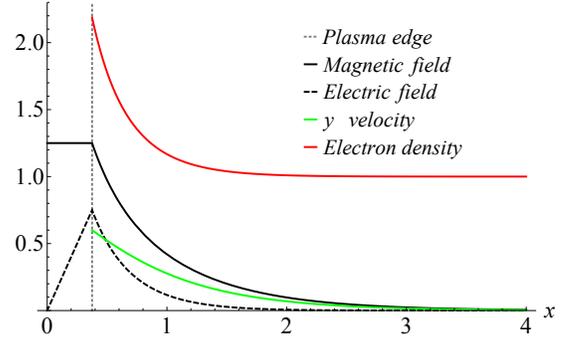

FIG. 4. The approximate solution in the "broken mirror" scenario in the case $\tilde{\gamma}_0 = 1.25$, $B_0 = 0$. The black solid line is the magnetic field $B(x)$, the black dashed is the electric field $E_x(x)$, the red line is electron density $n(x)$ and the green line is the y-component of electron velocity $v_y(x)$. The x-component of the electron velocity $v_x(x) = 0$ in this case.

The components of the Maxwell's equations, which are not identically zero, expressed through the currents and densities, are

$$\left. \mathrm{rot}\vec{B}\right|_y = -\frac{4\pi e}{c}\left(n_{e\alpha}(x) v_{y\alpha}(x) + n_{e\beta}(x) v_{y\beta}(x)\right), \tag{6a}$$

where, according to figure 1, $v_{y\alpha}(x) > 0$, and $v_{y\beta}(x) < 0$;

$$\mathrm{div}\vec{E} = 4\pi e\left(2 n_i - n_\alpha(x) - n_\beta(x)\right) \tag{6b}$$

Denoting $v_0$ as the absolute value of the flow velocity at $\pm\infty$, and $B_{max}$ as the maximum value of the magnetic field, which evidently corresponds to the central region between plasmas, it is possible to introduce characteristic values of the problem as

$$[\text{time}] = \omega_e^{-1}, \tag{7a}$$
$$[\text{length}] = c/\omega_e, \tag{7b}$$
$$[\text{fields}] = m_e c\,\omega_e/e, \tag{7c}$$
$$[\text{velocity}] = c, \tag{7d}$$
$$[\text{density}] = n_0, \tag{7e}$$

where

$$\omega_e \equiv \sqrt{\frac{4\pi n_0 e^2}{m_e}}. \tag{8}$$

Now, taking into account the symmetry $(x,y) \to (-x,-y)$, which allows to consider only one flow, in the dimensionless units the system of equations (below we make all the calculations for one flow with the positive $v_x$ and $x$, solutions for the other flows are evident) reads

$$\frac{dnv_x}{dx} = 0 \tag{9a}$$

$$\frac{dB}{dx} = 2nv_y, \tag{9b}$$

$$\frac{dE_x}{dx} = 2(1 - n), \tag{9c}$$

$$v_x \frac{d}{dx} \gamma v_x = -E_x - v_y B, \tag{9d}$$

$$v_x \frac{d}{dx} \gamma v_y = v_x B - E_0, \tag{9e}$$

where we introduced a constant electric field $E_0$ along the $y-$axis, which will be defined from the boundary conditions below. The parameters, which define the solution of (9), are the initial velocity $v_0$, which is the boundary condition for the absolute value of $v_x$ in (9) at $x \to \pm\infty$, and the magnetic field $B_0 = B(x \to \pm\infty)$, which is related to the electron magnetization $\sigma = ceB_0/m_e v_0 \omega_e$ in astrophysics[2,3,10].

### B.  Compressed magnetic mirror model: Reflecting mirror

The solution of the system (9) may be simplified in the case where the magnetic field dominates in the Lorentz force $v_0 B \gg E_x$, and the electric field $E_x$ can be neglected. One possibility may be realized if some electrons are present in the region between colliding plasmas and partially neutralize ions there. This may happen in some regimes of the plasma interpenetration. Another possibility to neglect the electric field may be realized in the case of electron-positron flows, or, for electron-ion flows, in the ultrarelativistic limit $v_0 \to 1$.

When the magnetic force dominates, in the first approximation $E_x$ may be omitted in (9d). Then, using (9e) to substitute $B$ with $v_y$, one obtains the energy conservation law

$$\gamma + \frac{BE_0}{2v_0} = const. \quad (10)$$

Equation (9a) means $nv_x = v_0$. From (10) we get the function $v_x(v_y)$. Multiplying equation (9d) by $n$ and taking into account (9a) and (9b),

$$B(v_x) = \sqrt{B_0^2 + 4v_0(\gamma_0 v_0 - \gamma v_x)}, \quad (11)$$

where $\gamma_0 = 1/\sqrt{1-v_0^2}$. From (11) the magnetic field in the sheath $|x| < x_0$ void of electrons is

$$B_{max} = \sqrt{B_0^2 + 4\gamma_0 v_0^2} \quad (12)$$

for magnetized plasmas with $B(x \to \pm\infty) = B(v_x \to v_0) = B_0$. Equation (11) provides an implicit expression for the magnetic field as the relativistic factor $\gamma$ is related to $B$ according to (10). Expressing $v_y$ and $\gamma$ as functions of $v_x$, and using (9), (10), and (11), one can find

$$v_y(v_x) = -\sqrt{1 - v_x^2 - \frac{1}{\gamma^2(v_x)}}, \quad (13)$$

and

$$B(v_x) = v_x E_0 + \sqrt{(B_0 - v_x E_0)^2 + 4v_0\gamma_0(v_0 - v_x)}, \quad (14)$$

The condition $v_y(x \to \infty) = 0$ defines $E_0 = v_0 B_0$. An equation for $x(v_x)$ that follows from (9e)

$$dx = \frac{d}{dv_x}(\gamma(v_x)v_y(v_x)) \frac{v_x \, dv_x}{v_x B(v_x) - E_0}, \quad (15)$$

may be numerically integrated to obtain explicit dependencies $B(x)$, $v_x(x)$, $v_y(x)$ and $n(x)$. The integration contains a constant $x_0$, which corresponds to the thickness of the magnetic sheath and may be defined from the condition (12) and the conservation of the magnetic flux. It depends on the preceeding history of the interaction.

In a compact form equation (15) can be integrated for $B_0 = 0$:

$$x(v_x) - x_0 = \sqrt{\gamma_0}\left(1 - \sqrt{1 + \frac{v_x}{v_0}} - \right.$$
$$\left. -\frac{1}{2\sqrt{2}}\ln\left((3 + 2\sqrt{2})\frac{\sqrt{2} - \sqrt{1 - v_x/v_0}}{\sqrt{2} + \sqrt{1 - v_x/v_0}}\right)\right) \quad (16)$$

Then one has to find the inverse function $v_x(x)$, which is a numerical procedure since the function (16) is transcendental. With the function $v_x(x)$, or, using (10), $v_y(x)$, for which $E_x(v_y)$ looks more compact, density $n(x)$ and electric field $E_x(x)$ may be found from (9a) and (9c) correspondingly. Density is then expressed as $v_0/v_x$, and, as functions of $v_y$, the electric field may be found by integration of (9c). In case $B_0 = 0$, which is a good approximation, if $B_{max} \gg B_0$,

$$E_x(v_y) = 2\sqrt{\gamma_0}\left(\sqrt{2} - \frac{v_y/v_0}{\sqrt{1 - \sqrt{1 - v_y^2/v_0^2}}}\right) \quad (17a)$$

$$B(v_y) = 2v_0\sqrt{\gamma_0}\sqrt{1 - \sqrt{1 - v_y^2/v_0^2}}. \quad (17b)$$

The stationary solution for $v_0 = 0.9$ is shown in panel I of figure 2. As expected, the magnetic field (black solid line), is greater than the electric field (black dashed line) for not small $v_0$. However, for $v_0 \to 0$ it turns $E_x \gg v_0 B$, which is reasonable, since the magnetic part of the Lorentz force is small for small velocities. It is seen from figure 2, panel II, that for high enough $v_0$ the magnetic field is everywhere larger, than $E_x$. This allows us to use the considered approximation even in case of an uncompensated ion electric charge in the region of the magnetic field between electron flows. It also shows, that, at least for large enough $v_0$, the iterational procedure may be developed to find a more precise solution of (9).

In the case of a magnetized electron flux the solutions (13) and (14) with the integration of (15) can not actually be a stationary solution for both $\alpha$ and $\beta$ fluxes, because in the plasma region, the constant $E_0$ should be of the opposite signs for opposite signs of $v_x$. However, we mention this solution here, as it may be used as an initial condition for the reflection problem, at the time moment, when the incident plasma flow stops.

Above the electron-ion plasmas were considered. It is possible, in the same conditions, to consider also electron-positron plasmas. The densities and velocities of electrons and positrons are $n^-$, $n^+$, $\vec{v}^-$ and $\vec{v}^+$ respectively. As the masses of both species are equal and the signs are opposite, the stationary solution should satisfy the following relations

$$n^+ = n^-, \quad (18a)$$
$$v_x^+ = v_x^-, \quad (18b)$$
$$v_y^+ = -v_y^-, \quad (18c)$$
$$E_x = 0. \quad (18d)$$

As a result of the latter relation, the presented earlier approximate solution of (9) for the magnetic field (14), densities and velocities (16) becomes exact for an arbitrary $v_0$. Figure 2 shows the solution for one of the flows (say electrons, with positive velocities), it applies in this case for the densities $n^-(x)$, velocities $v_x^-(x)$ and $v_y^-(x)$, and magnetic field $B(x)$, but not for the electric field which is zero in this case, $E_x(x) = 0$.

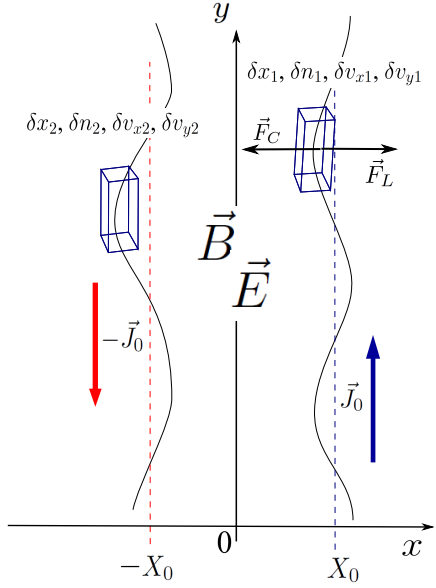

FIG. 5. The scheme for the linear analysis of the instability.

## C. Compressed magnetic mirror model: Broken mirror

### 1. Quasi-stationary hydrodynamic solution

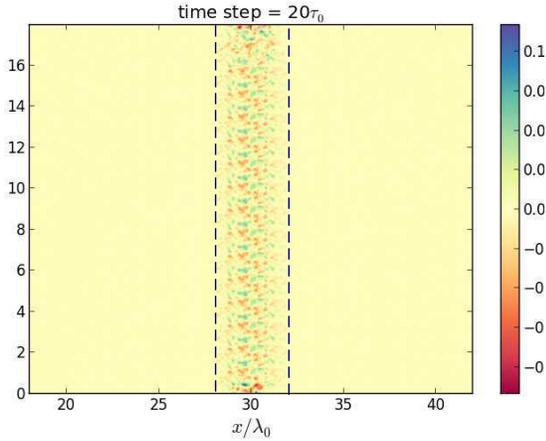

FIG. 6. The fluctuating magnetic field for two colliding electron-ion flows for the time moment of 20 plasma periods $2\pi/\omega_e$. Ambient magnetic field is absent. Dashed lines circumscribe the intersected ion flows.

The solution of the hydrodynamical model presented in the previous section, describes the electron reflection by a compressed magnetic field. However, one can suppose that in some situations, depending on the preceeding history of the interaction, the stationary solution may not be achieved. In the case of electron-ion plasmas, the energy of the electron flow may be partly redistributed into electron-ion interaction during the process of electron slowing down. As a result, the back-scattering of electrons with the same final velocities becomes impossible. The situation may correspond to electron flows, moving along the edge of the magnetic mirror perpendicularly to the ion flow and the magnetic field, i.e. when the electric force, created by ions $F_C$ and the magnetic part of the Lorentz force $F_L$ are near the equilibrium. The corresponding condition is given by:

$$E_x = -v_y B. \quad (19)$$

Let us consider equation (9) in the limit $v_x \to 0$. Equation (19) then follows from (9d). Taking differentiation of (19), and using (9b) and (9c), it is possible to find the density

$$n = \frac{B(B-B_0)}{2\gamma} + \gamma^2, \quad (20)$$

where, in order to resolve the ratio $E_0/v_x$ in the limit of $v_x \to 0$ and $E_0 \to 0$, we applied a condition $B_0 v_x = 2E_0$, which gives the correct limit value for the density $n(x \to \infty) \to 1$. From (9e) then

$$\frac{dv_y}{dx} = \frac{1}{\gamma^3}\left(B - \frac{B_0}{2}\right). \quad (21)$$

The magnetic field $B(v_y)$ from (9b) is

$$B(v_y) = \frac{B_0}{2} + \sqrt{\frac{B_0^2}{4} + 4\gamma^2(\gamma - 1)}, \quad (22)$$

and for the definition of velocity $v_y(x) = \sqrt{1 - 1/\gamma^2}$ from (21) there is an equation

$$\frac{\gamma}{\sqrt{\gamma^2-1}}\frac{d\gamma}{dx} = \sqrt{\frac{B_0^2}{4} + 2\gamma^2(\gamma-1)}. \quad (23)$$

The thickness of the non-neutral layer $x_0$ in (23) may be related to $B_{max}$ and the flow gamma factor $\tilde{\gamma}_0 \equiv 1/\sqrt{1-v_y^2(x_0)}$ by the condition of equilibrium (19) as $B_{max} = -E_x(x_0)/\sqrt{1-\tilde{\gamma}_0^{-2}}$. The electric field at the boundary is defined by the uncompensated ion charge as $E_x(x_0) = 2x_0$. Note, that now the maximum value of the magnetic field

$$B_{max} = \frac{B_0}{2} + \sqrt{\frac{B_0^2}{4} + 4\tilde{\gamma}_0^2(\tilde{\gamma}_0 - 1)}, \quad (24)$$

is related with the $\tilde{\gamma}_0$ and the sheath layer thickness $x_0$: the greater is $\tilde{\gamma}_0$, the greater is $x_0$, the greater is $B_{max}$.

Let us consider an unmagnetized flow $B_0 = 0$. The equation for $\gamma(v_y)$ is

$$x_0 - x = \frac{1}{2\sqrt{2}}\ln\left[\frac{\sqrt{2}-\sqrt{1+\gamma}}{\sqrt{2}-\sqrt{1+\tilde{\gamma}_0}}\frac{\sqrt{2}+\sqrt{1+\gamma}}{\sqrt{2}+\sqrt{1+\tilde{\gamma}_0}}\right], \quad (25)$$

$$x_0 = \sqrt{\tilde{\gamma}_0 - 1}\sqrt{\tilde{\gamma}_0^2 - 1}. \quad (26)$$

Further, the magnetic and electric fields are defined from (22) and (19), where $B_0 = 0$. The electron density is defined by (20). Note, that it does not diverge at the plasma boundary. The results for the numerical procedure (25), and the corresponding fields are shown for $\tilde{\gamma}_0 = 1.25$ in figure 4.

According to (25), in the case $\tilde{\gamma}_0 - 1 \ll 1$ the characteristic width of the current sheet $x - x_0 \gtrsim 1$, and $x_0 \ll 1$. That is, the current sheet is much thicker than the width of the non-neutral zone. In the opposite case, $\tilde{\gamma}_0 \gtrsim 1$, from (25) it follows, $x - x_0 \sim 1/\sqrt{\tilde{\gamma}_0}$, and $x_0 \sim \tilde{\gamma}_0^{3/2} \gg x - x_0$. The later relation means, that the plasma flow is much thinner than the non-neutral zone and it may be approximated with a thin current

$$J_0 = 2 \int_{\tilde{\gamma}_0}^{1} (n(\gamma) - 1) v_y(\gamma) \frac{dx}{d\gamma} d\gamma =$$
$$= 2\sqrt{2} \left( \tilde{\gamma}_0 \sqrt{\tilde{\gamma}_0 - 1} - \arctan \sqrt{\tilde{\gamma}_0 - 1} \right), \quad (27)$$

created by electrons of the average density

$$N_0 = 2 \int_{\tilde{\gamma}_0}^{1} (n(\gamma) - 1) \frac{dx}{d\gamma} d\gamma = 2\sqrt{2}(\tilde{\gamma}_0 - 1)\sqrt{\tilde{\gamma}_0 + 1}, \quad (28)$$

moving with the average velocity

$$U_0 = J_0/N_0, \quad (29)$$

along the edges

$$\pm X_0 = \pm x(\Gamma_0), \quad (30)$$

where $\Gamma_0 \equiv \sqrt{1 - U_0^2}$.

The model of thin edge currents with parameters (27), (28), (29), and (30) is used below for the stability analysis in section II C 2.

Note, that this type of solution can not be truly stationary. It can be only an intermediate step in some sequence of quasi-equilibrium stages of the system, since the upcoming electron flux changes the plasma parameters near the edge with time. The stability of this solution is analyzed in the next section.

### 2. Edge instability

In this section we show, how and why the electron-ion plasmas can break the "magnetic mirror". As it is mentioned above, and also seen from PIC modelling (see below) at the beginning of the interaction when the compressed magnetic field between two plasmas stops electrons, the density of electron flows is increased substantially near the plasma edge. When this happens, the sharp plasma sheets are formed by electrons, which at this moment are near the equilibrium (19). To analyze the stability of the edge, we consider the following model: two infinitely thin sheets of currents separated with the magnetic field barrier (see figure 5). These sheets consist of electrons in electron-ion flows, and in this case the constant ion background is everywhere, so there is also nonzero electric field between the current sheets.

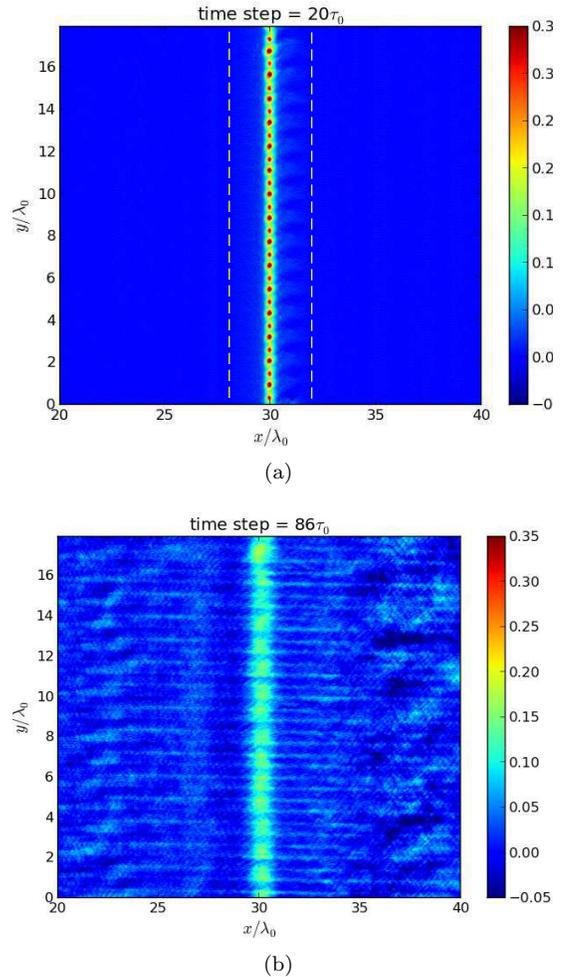

FIG. 7. Magnetic field at the time moments of 20 (a) and 86 (b) plasma periods for the flows interpenetrating in the moderate external magnetic field $B_0 = 0.016$ (5 T).

In figure 5 dashed lines at $x = \pm X_0$ correspond to the initial positions of $\delta$-function currents $\pm J_0 = \mp N_0 U_0$. The model is defined by one parameter $\tilde{\gamma}_0$ according to (27), (28), (29) and (30), their values depend on the history of the plasmas collision. Fields $\vec{E}$ and $\vec{B}$ acting on an electron are equilibrated for these initial positions of currents according to (19).

Let us consider the small change in position, velocities, densities $\delta x_1$, $\delta v_{x1}$, $\delta U_1$, $\delta N_1$, $\delta x_2$, $\delta v_{x2}$, $\delta U_2$, $\delta N_2$ in the current sheets #1 and #2. These fluctuations change both electric and magnetic fields, and if the fields act so that the initial small fluctuation increases, then the system is unstable. According to this, the instability on the edge is analysed as follows: (i) Assume the small changes of all physical values in currents and densities; (ii) find the change in fields $\delta \vec{B}$ and $\delta \vec{E}$; (iii) calculate the action of the perturbed fields on the currents and densities. The details of calculations are presented in

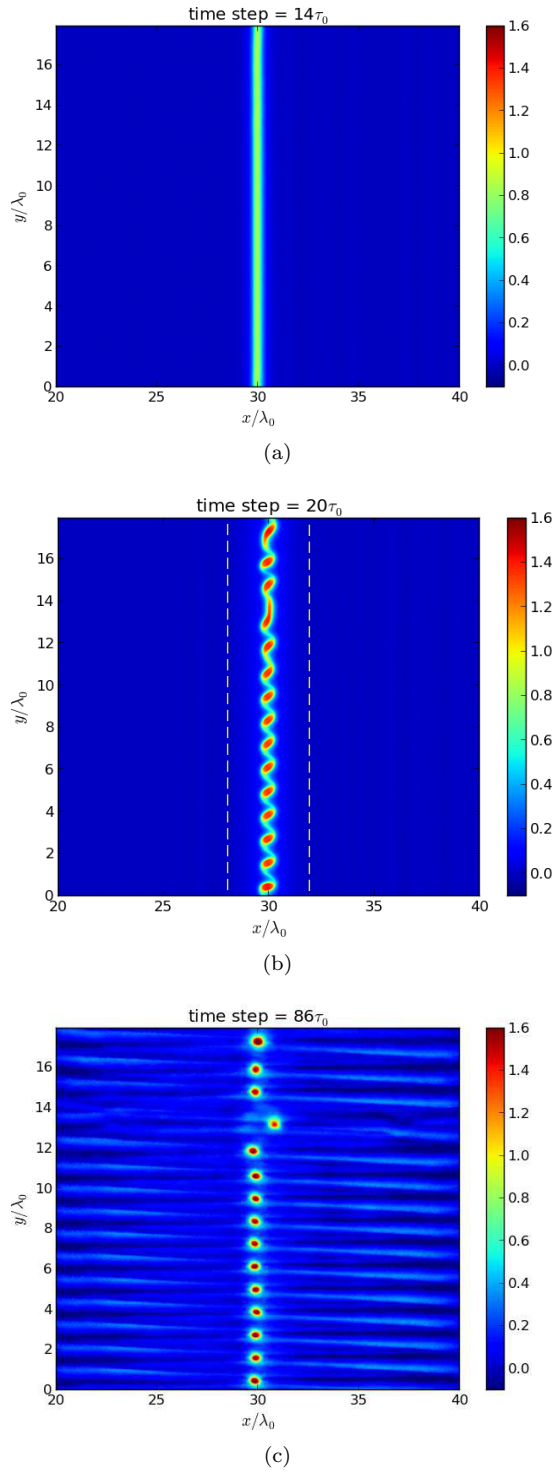

FIG. 8. Magnetic field at the time moments of 14 (a), 20 (b) and 86 (c) plasma periods for the flows interpenetrating in the external magnetic field $B_0 = 0.078$ (25 T). Dashed lines circumscribe the intersected ion flows.

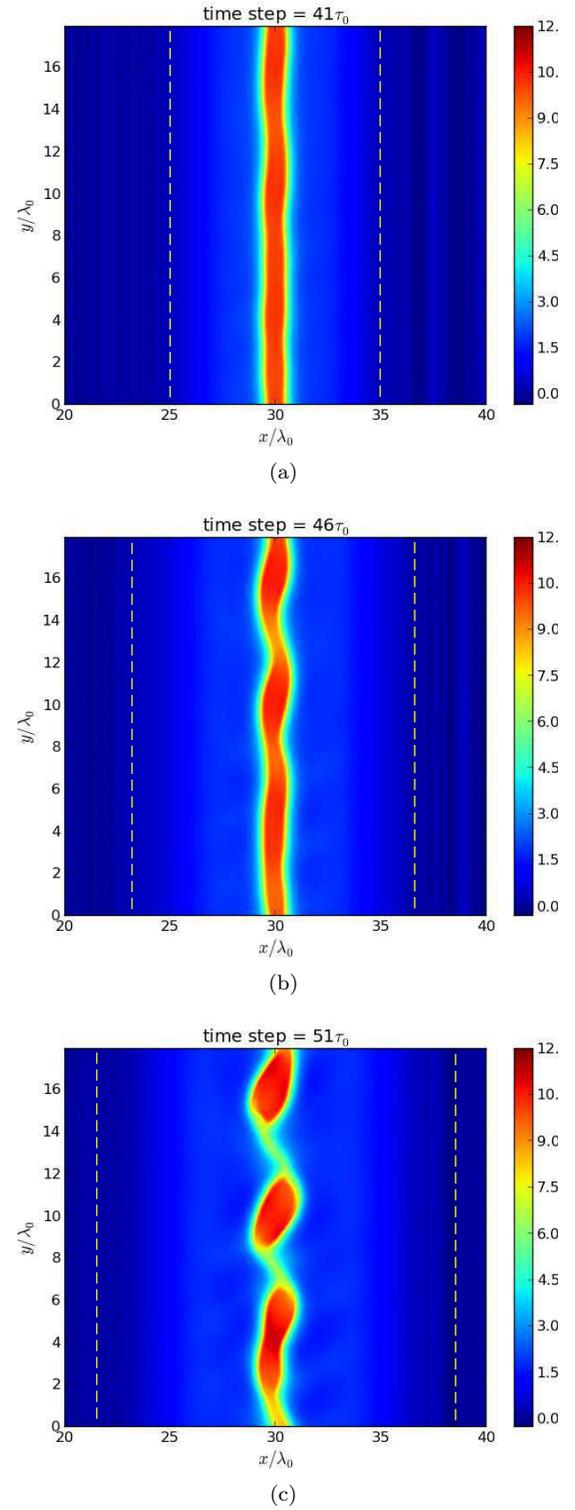

FIG. 9. Magnetic field at the time moments of 41 (a), 46 (b) and 51 (c) plasma periods for the flows interpenetrating in the external magnetic field $B_0 = 1.56$ (500 T). Dashed lines circumscribe the intersected ion flows.

Appendix.

As it follows from (A.11), in the first highest order of

$\tilde{\gamma}_0 \to \infty$ frequency $\omega(k)$ is defined by the equation

$$2\omega^4(\omega^2 - (1-Z)^2 k^2) + (\omega^2 - k^2 Z^2) Q^2 - \qquad (31)$$
$$-\sqrt{2}\omega^2 Q(\omega^2 - (Z^2 - Z)k^2)(\tanh QX_0 + \coth QX_0) = 0,$$

where $Q = \sqrt{k^2 - \omega^2}$ and $Z$ are defined by (A.5d). There are several possible branches. The two lowest branches are $\omega^2 = k^2$ and $\omega^2 = Z^2 k^2$, they correspond to a constant group velocity. There is also a series of branches, which correspond to the waveguide modes, starting at the solutions $\omega_n$ of (31) for complex $Q$, and having the form of $\omega(k) \approx \omega_n \pm k^2$ for small $k$.

The real branches $\omega^2 = k^2$ and $\omega^2 = Z^2 k^2$ merge at some $k^*(\tilde{\gamma}_0)$, which results in the complex solutions for $k > k^*(\tilde{\gamma}_0)$. However, these solutions are not those ones which we are looking for, since the value $k^*(\tilde{\gamma}_0)$ increases with the increasing of $\tilde{\gamma}_0$, which does not agree with the results of numerical simulations and the qualitative expectations. In the limit $\tilde{\gamma}_0 \to \infty$ corrections to the main terms, coming from taking into account other terms in (A.11) do not give the qualitative changes and complex values for $\omega(k)$. The reason of this comes from the properties of Eq. (A.11), where the highest order terms cancel in the limit $\tilde{\gamma}_0 \to \infty$.

A more accurate analysis of the dispersion equation (A.11), using the condition $\tilde{\gamma}_0 \gg 1$, gives the following result. The two branches $\omega^2 = k^2$ and $\omega^2 = Z^2 k^2$ interact also in the region of long wavelengths $k \lesssim \tilde{\gamma}_0^{-3/2}$, giving complex solutions to $\omega(k)$. The approximate dispersion relation in this region with the accuracy $\sim k^4 \sim \omega^4$ reads

$$\Upsilon(B_0 - (1-Z)N_0) + \Gamma_0^3[p_+\beta_-(k-\omega)^2 + p_-\beta_+(k+\omega)^2] -$$
$$-\sqrt{2}\omega^2 N_0[p_+(1-\beta_-) + p_-(1-\beta_+)] = 0, \qquad (32)$$

where

$$\Upsilon \approx 2Q^2 + 2\omega^4(1-\beta_-)(1-\beta_+) -$$
$$-\sqrt{2}\omega^2(2-\beta_+-\beta_-)(X_0^{-1} + 4Q^2 X_0/3),$$
$$p_\pm \approx X_0^{-1} + 4Q^2 X_0/3 - 2\sqrt{2}\omega^2(1-\beta_\pm),$$
$$\beta_\pm \approx k/(Zk \pm \omega).$$

The instability condition can be obtained from the analysis of the group velocity $\alpha = \partial\omega(k)/\partial k|_{k=0}$ in the limit $k \to 0$. The real and imaginary parts of $\alpha$ as functions of $\tilde{\gamma}_0$ are shown in figure 10. There are four regions of $\tilde{\gamma}_0$. In the region $\tilde{\gamma}_0 < \tilde{\gamma}_1 \approx 7.2$, there are four complex roots, with real and imaginary parts. Then, for $\tilde{\gamma}_1 < \tilde{\gamma}_0 < \tilde{\gamma}_2 \approx 8$, the only pure imaginary solutions are possible. In the region $\tilde{\gamma}_2 < \tilde{\gamma}_0 < \tilde{\gamma}_3 \approx 11$, there are two real and two imaginary roots, and finally, for $\tilde{\gamma}_0 > \tilde{\gamma}_3$, there are no complex roots. The latter domain corresponds to the limit $\tilde{\gamma}_0 \to \infty$, where Eq. (31) applies.

The numerical simulations presented in the next section were conducted in the domain $\tilde{\gamma}_0 < \tilde{\gamma}_1$. The observed instability is in agreement with the present analysis. The wavelength of the unstable mode is of the order of the sheath thickness, $k \lesssim \tilde{\gamma}_0^{-3/2}$, and the growth rate,

$\Im\omega \sim \tilde{\gamma}_0^{-3/2}$, decreases with increasing the magnetic field strength which is proportional to $\tilde{\gamma}_0^{3/2}$, according to Eq. (24).

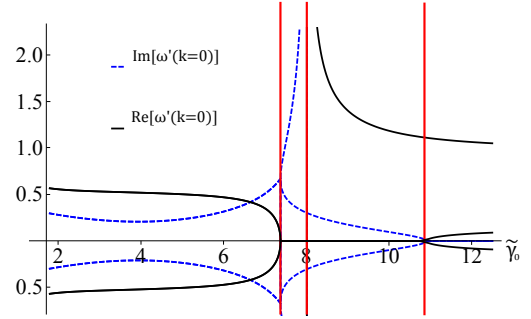

FIG. 10. The real $\Re[\omega'(k=0)]$ and imaginary $\Im[\omega'(k=0)]$ parts of the wave group velocity $\partial\omega(k)/\partial k|_{k=0}$ as a function of $\tilde{\gamma}_0$ for the solutions of the dispersion relation (32).

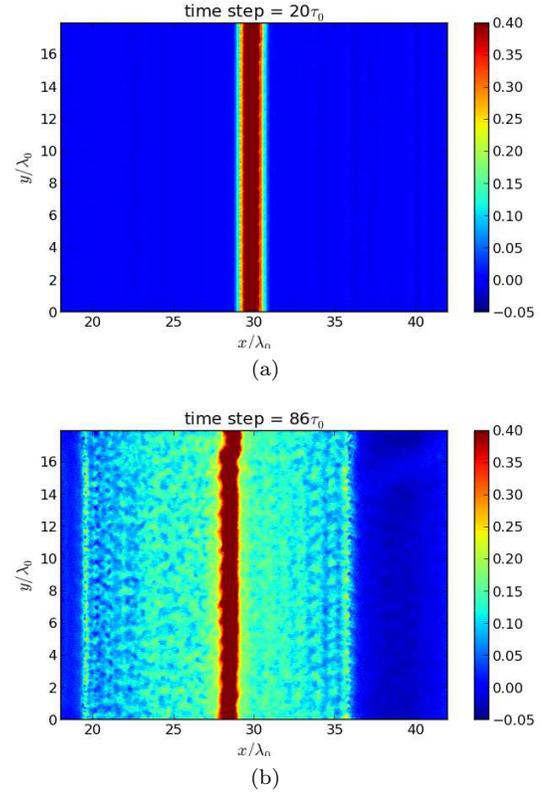

FIG. 11. Magnetic field at the time moments 20 (a) and 86 (b) plasma periods for the electron-positron flows interpenetrating in the external magnetic field $B_0 = 0.078$ (25 T).

### III. NUMERICAL SIMULATIONS

Numerical simulations of the interpenetration of two collisionless plasma flows are compared with the simple models, described above. The ambient magnetic field is modeled by introducing the proper magnetic component in the Lorentz force acting on plasma particles. This approach neglects the preceeding history of plasma interaction with the ambient magnetic field, and defines the initial stage as a magnetized plasma in a homogeneous ambient magnetic field.

For particle-in-cell simulations we used PICLS code[19], modified in order to treat the external magnetic field. In all simulations presented below the parameters of both flows are equal, if the opposite is not explicitly stated. Initial densities of electrons, ions and positrons in each flow are $10^{18}$cm$^{-3}$, the temperatures of all the species are 100 eV, and the flow velocities are $0.2c$. These parameters, which correspond to a typical laboratory experiment, can be rescaled to any other plasma density according to the relations (7). In the figures the length is normalized by $\lambda_0 = 2\pi c/\omega_e$ and the time – by $\tau_0 = 2\pi/\omega_e$. Periodic boundary conditions were used in the perpendicular (y) direction. The temporal resolution in simulation was 40 calculation steps per plasma period, and the number of quasi-particles was 40 per cell. The initial distance between plasmas was 200 $\mu$m, or $\approx 6 \times 2\pi c/\omega_e$. To avoid numerical artefacts, the initial plasma profiles were smoothed at the front and back edges.

The results of the simulations are presented in figures 6, 7, 8, 9 and 11, where the magnetic field $B_z(x,y)$ is plotted for different initial conditions and at different time moments. The geometry of the interaction in figures 6, 7, 8, 9 and 11 is the same as in figures 1, 3, and 5. The full movies, which show the time-evolution of $B_z(x,y)$ and the electron density $n(x,y)$ are presented as supplementary multimedia files.

### A. Electron-ion plasmas

Firstly, the simulation results for plasmas collisions without an external magnetic field are shown in figure 6 at the time moment $20 \times 2\pi/\omega_e$. At this time plasma flows are already interpenetrated to each other. The magnetic field shown in the figure, is generated as a result of the electromagnetic (Weibel) electron-electron instability[20].

When a comparatively small ambient magnetic field $B_0 = 0.015$ (5 T) corresponding to a rather weak electron magnetization $\sigma = 0.08$ is on, at the same time moment $20 \times 2\pi/\omega_e$ (see figure 7(a)) the magnetic field configuration is completely different. It is seen, that in the overlapping region the value of magnetic field is much greater, than the fluctuations in figure 6. Moreover, its maximum value, which is of the order of 0.35, exceeds more than 20 times the initial external magnetic field. This value is achieved by squeezing the initial magnetic field by a factor about 20, so the size of magnetic structures should be $\sim 0.2..0.3\lambda_0$, in accordance with figure 7(a). This size, as well as a value of the maximum magnetic field in the comb, may be compared with either (12) or (24). However, it follows from the geometry of plasma flows, that the stationary solution (14), which corresponds to the case B in figure 1, is not realized. What is seen in figure 7(a), is the developed in time quasi-stationary solution (22), broken by the edge instability, described in section II C 2.

Magnetic field, which is organized as a comb of separated regular islands with the size corresponding to the estimate (30), allows electrons to penetrate through it, though there is no electrons inside the islands of magnetic field. The modulation period is of the order of the sheath thickness. The magnetic structure works like a comb for electron flows, introducing the pertrubation and additional electron heating. Note that ions are not disturbed by the fields on the considered time scales, and everywhere, even inside the magnetic field island they represent a constant positively charged background.

Further in time the comb is degraded, and the electrons begin to penetrate inside magnetic field islands, as it is seen in figure 7(b). This process takes time, and for higher magnetic fields, considered below, the time is so large, that it goes beyond the time scale considered in the present paper. While the intensity of the external mag-netic field increases, the structure becomes more resolved and more rigid, which is seen in figures 8, representing the case of a stronger electron magnetization with the initial external magnetic field $B_0 = 0.078$, (25 T, corresponding magnetization is $\sigma \approx 0.4$). Initially much lower, the magnetic field is compressed by the merging plasmas, up to the value, which is about 20 times greater than the initial one. In figure 8(a) the field at the time moment $14 \times 2\pi/\omega_e$ is shown, and it reminds the configuration from figure 1 or 3. At this moment the electron flows do not penetrate through the compressed magnetic field (see multimedia files for details). Again, as in the case of the lower external field $B_0 = 0.015$ one may expect the reflection of electron fluxes, however, what happens in reality, is the destruction of the magnetic mirror, shown in progress in figure 8(b). This process is slower than for $B_0 = 5$ T, and the scale of the comb length is greater. The quasi-stationary solution (22) allows to correctly estimate the scale of $B_{max} \approx 1.6$, $x_0 \approx 0.5$, and the maximum electron density on the edge of the plasmas $n_{max} \approx 3$.

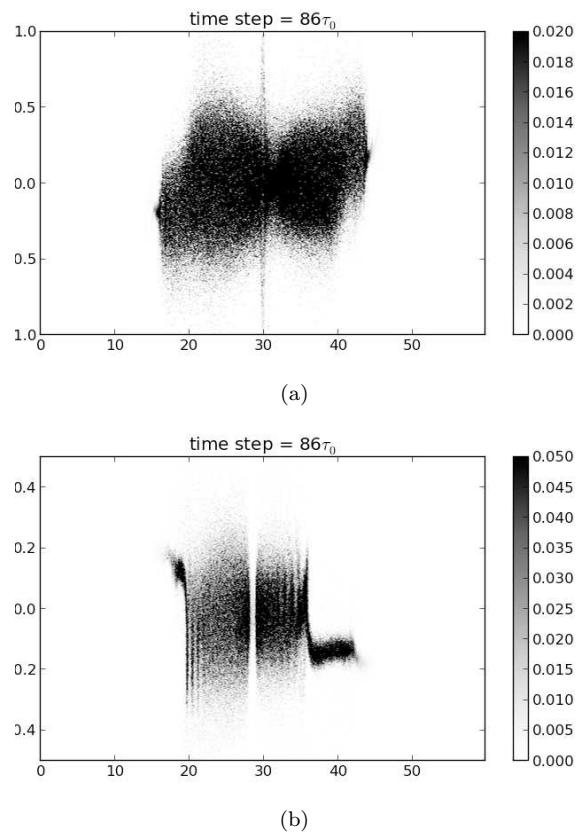

FIG. 12. The phase plots at the late time of 86 plasma periods shows the heating of electrons in electron-ion (a) and electron-positron (b) flows in the external magnetic field $B_0 = 25$ T.

Later the self-organized "comb" cut electron flows and produces electron regular structures, as shown in figure 8(c). At the considered time scale this structure does not develop enough to influence ion motion and the total flow behavior, and the detailed studies of this structures will be presented elsewhere.

Our primary interest is the transition from magnetic mirror to "comb". For better understanding, we simulated plasmas collisions for very high initial magnetic field[21] $B_0 = 1.56$ corresponding to strongly magnetized electrons ($\sigma \sim 0.8$) in the field of 500 T (see figure 9 and multimedia files for more detailed description). In figure 9 the transition from the magnetic mirror to the magnetic "comb" is shown in three steps. The first step, figure 9(a), at the time moment $41 \times 2\pi/\omega_e$, the magnetic field is compressed by a factor of 10 but it is not yet a "broken mirror", only small deviations from the straight line can be seen at the edges. The second step, figure 9(b), at the time moment $46 \times 2\pi/\omega_e$, corresponds to the intermediate situation, when the magnetic field does not grow any more and the two edges are still separated, but the comb structure is already developing. Finally, after the third step, figure 9(c), at the time moment $51 \times 2\pi/\omega_e$, magnetic field is separated in vortices and cut electron flows, which now are able to cross the border.

In the case of strong $B_0 \sim 1$, from the "broken mirror scenario", both the maximum values and the gradients of the electron density $n(x)$ and electron velocity $v_y(x)$ appear to be very high. The resolution of PIC does not allow to go up to the predicted values, and the comparison fails. It is probable, that in the case $B_0 \sim 1$ the considered above simple analytical models should be reconsidered, taking into account the instability developing before the forming of the quasi-stationary solution, the perturbation of ion density, the magnetic field diffusion, etc.

As we see from the results of PIC simulations, a stationary reflection of electron flows from the magnetic barrier, as predicted by the simple "compressed magnetic mirror" model, is hardly possible. The explanations of this fact consists of two points. The first one is that the approximation, made in section II B concerning a dominant role of magnetic field in the electron Lorentz force is not valid. The second one is the instability, analyzed in section II C 2, which develops at the edges of the electron sheath and breaks the magnetic "mirror" before electron flows start to reflect.

### B. Electron-positron plasmas

In the case of electron-positron flows the situation changes dramatically. The simulations were performed for a relatively weak initial magnetization $\sigma \approx 0.4$, corresponding to the initial ambient magnetic field $B_0 = 0.078$ or 25 T. As it is shown in figure 11, the magnetic field compression is weaker and the magnetic mirror is much more stable, than in case of electron-ion flows. This is evident already at the time moment $20 \times 2\pi/\omega_e$ (compare figures 11(a) and 8(b)). Further, as it is shown in figure 11(b), electrons and positrons, are reflected from the magnetic field. This reflection reminds the initial stage of the magnetic reflection dominated shocks in the limit of strong magnetic fields.

As it was shown in section II B, the solution corresponding to hydrodynamical reflection is exact in the case of electrons and positrons, and it is valid for an arbitrary $v_0$. At the same time, the instability, which slowly deforms the "magnetic mirror" is very much different. From the solution (17b), suitable for the reflection by a "magnetic mirror" model, it is possible to estimate the maximum value of the magnetic field. For the simulation value of $v_0 = 0.2$, from (12) $B_{max} \approx 0.4$, which coinsides with the maximum value of the magnetic field in figure 11.

It is interesting to compare the phase space for electrons in electron-ion and electron-positron flows. In the case of electron-ion flows (figure 12(a)), there is a trapped fraction of electrons in the magnetic mirror, at $x = 30$. These are electrons, rotating near the magnetic tubes. The electrons are heated on the considered time scale presumably in the ion front regions, but also in the magnetic barrier. This electron heating is more efficient, than that for electron-positron flows (figures 12(a) and 12(b)). On the later case, in figure 12(b), it is clearly seen, that electrons can not penetrate through the border, formed by the compressed magnetic field, the reflection is evident. The heating, seen in figure 12(a) takes place at the plasma fronts and it deserves a separate study. This is likely the electron-type instability, which has a characteristic time of the order of $\omega_e^{-1}$.

### IV. DISCUSSION

The experimental examination of the proposed theoretical description of the collisionless electron-ion plas-mas interaction is in principle possible with the modern laser facilities. The time of life of the magnetic field should be of the nanosecond order considering the plasma density $\sim 10^{18}$ cm$^{-3}$ and the velocity $\sim 0.2c$, and the distance between plasmas at the beginning of interaction – of the order of a few hundred of microns. Then the effects at the time scale of a few hundred of electron times $\tau_0$ may be observed. It is a question if it is possible within these distance, time, and energy scale to go beyond the electron time scale and to catch some ion physics. The presented theoretical study shows, that it may be difficult since we did not observe a significant change of the ion distribution function in PIC simulations for the 25 T magnetic field.

The specificity of laboratory experiments with laser produced plasmas is that the plasma life time is comparable or less than the collisionless shock formation time. This defines certain difficulties in the comparison between observations and laboratory experiments. The initial stage of interaction, which is the main object of the present study, is the crucial element for understanding of the further development of shocks. A compression of the magnetic field in case of electron-ion plasmas may lead to long-living regular structures, which cause a significant plasma heating and may affect the physics of shock formation. The even more pronounced effect is seen in case of electron-positron plasmas. In this case, the magnetic barrier is not breaking for the observed simulation times and works like a mirror, reflecting electron-positron plasmas back.

Application of the considered transient model to astrophysical conditions needs further studies. Our hypothesis of the external magnetic field between colliding plasmas may be oversimplifying, a third plasma could be considered, which however, may be of a much lower density, but with a stronger magnetization. A transition between the collisional and collisionless shocks may be also considered. A shock may be formed as a collisional shock as, for example, in a supernova remnant and then be transformed in the collisionless shock as the density decreases and the magnetic field increases.

As it was seen in simulations, the electron-ion and electron-positron plasmas behave in a different way in the considered interaction geometry. The first one may be described in terms of the "broken mirror" model, while the latest – in terms of the "reflecting mirror" one. In the case of electron-ion plasmas in the considered geometry and initial conditions it is impossible to avoid the deposition of some kinetic electron energy into electron-ion interaction. The process starts when the compressed magnetic field begins to deviate electron motion, retarding them with respect to ions. The sufficient part of the electron kinetic energy is converted to the charge-separation electrostatic potential, which does not allow electrons to escape. They form the quasi-stationary flow along the magnetic barrier. Next transient state may be described with the linear instability analysis. The characteristic spacial properties of the resulting magnetic comb may be estimated by the velocity on the plasma edge, or by the barrier width $x_0$, which may be estimated from the model, presented in section II C 1. However, it is worth to mention here, that the estimates were made for cold collisionless plasma flows. Finite temperatures, as well as an energy transition to heating, may smooth plasma profiles and affect the stability analysis. After breaking of the "magnetic mirror", the energy redistribution leads to the further electron heating and evolution of the magnetic barrier which may end up with the ion trapping and shock formation.

In the case of electron-positron plasmas, there is no such energy redistribution, as in the electron-ion case, and, as a result, the reflection occurs almost elastically. This total reflection, which follows from a solution of the hydrodynamic equations (9), may, however, be broken down if the electrons and positrons are heated to high energies.

Here the simulations are limited to plasmas with relatively sharp edges. With a highly smoothed-edged plasmas profile the compressed magnetic field may penetrate into the region of a low plasma density. Then two colliding plasmas can partly overlap before they will be stopped by the magnetic field. At such special initial conditions, given as an example, the compensation of the ion charge by electrons in the region of the compressed magnetic field may occur, and the "reflecting mirror" scenario would become possible.

The scenario of the shock formation in the considered problem depends on the evolution of the magnetic barrier. Its width, height, and stability strongly depend on the initial conditions of the interaction, including initial magnetization, geometry, plasma composition, velocity, temperature (see Appendix), etc. The after-effect of the history of the shock formation on the shock structure in Space and laboratory is the subject of the future studies. But it is already clear, that in astrophysical experiments it is of a great importance to consider the interaction of the plasmas with the magnetic field in vacuum, and to take this interaction into account in the description of shock formation.

### ACKNOWLEDGMENTS

We wish to acknowledge the support of the French National funding agency ANR within the project SILAMPA and I. Andriyash for technical help. The numerical simulations were partially performed of the Department of the Theoretical Nuclear Physics of NRNU MEPhI.

### Appendix: Dispersion relation for the model of two current sheets in the magnetic mirror model

For the analysis of stability of the stationary solution we consider the model of two thin current layers, as shown in figure 5. The stationary fields at layer po-sitions are the magnetic $B_0 = 2\Gamma_0\sqrt{\Gamma_0 - 1}$ field and the electric field $E_0 = U_0 B_0$. The layer position $X_0$, the electron density $N_0$, the current $J_0$ and the velocity $U_0$ are defined by Eqs. (30), (27), (28) and (29), respectively.

Let us consider fluctuations of the denstites $\delta N_1$ and $\delta N_2$, the positions of the layers $\delta x_1$ and $\delta x_2$, the x-components of the velocities $\delta v_{x1}$ and $\delta v_{x2}$ and the y-components of the velocities $\delta U_1$ and $\delta U_2$. These are related to the fluctuations of electromagnetic fields $\delta B_z(x)$, $\delta E_x(x)$ and $\delta E_y(x)$, which obey the Maxwell's equations. In a linear analysis, all the fluctuations depend on the time and the position along the $y$ axis as $e^{-i\omega t + iky}$, and from (4) it follows, that in the inner region where there are no electrons

$$\delta E_x(x) = \frac{k}{\omega}\delta B_z(x), \quad (A.1a)$$

$$\delta E_y(x) = -\frac{i}{\omega}\frac{d\delta B_z(x)}{dx}, \quad (A.1b)$$

$$\frac{d^2\delta B_z(x)}{dx^2} - (\omega^2 - k^2)\delta B_z(x) = 0. \quad (A.1c)$$

Solutions of (A.1), in the inner region, are expressed with two constants $A$ and $C$ as

$$\delta E_x^{(in)}(x) = -kA/\omega \sinh Qx - kC/\omega \cosh Qx, \quad (A.2a)$$

$$\delta E_y^{(in)}(x) = -iQC/\omega \sinh Qx - iQA/\omega \cosh Qx, \quad (A.2b)$$

$$\delta B_z^{(in)}(x) = A\sinh Qx + C\cosh Qx, \quad (A.2c)$$

where

$$Q \equiv \sqrt{k^2 - \omega^2}. \quad (A.2d)$$

In the outer region with the equal densities of electrons and ions the evanescent fields can be written as

$$\delta E_x^{(out)}(x) = F_\pm \exp[\mp r_D^{-1}(x \mp X_0)], \quad (A.2e)$$

$$\delta E_y^{(out)}(x) = \mp i\sqrt{2}\omega D_\pm \exp[\mp\sqrt{2}(x \mp X_0)], \quad (A.2f)$$

$$\delta B_z^{(out)}(x) = D_\pm \exp[\mp\sqrt{2}(x \mp X_0)], \quad (A.2g)$$

where $F_\mp$ and $D_\mp$ are constants to be defined, $r_D$ is the plasma Debye length. Here we omitted in the linear approximation the electrostatic part of $\delta E_y^{(out)}$, and neglected a small electromagnetic part of the order of $\omega^2$ in $\delta E_x^{(out)}$. The smallness of $\omega \ll 1$ follows from the results as shown below.

The relations between the inner and outer solutions in (A.2) are defined by the fluctuations of the electron density, velocity and the position of the plasma layers:

$$\delta B_z\big|_{\pm X_0 - 0}^{\pm X_0 + 0} = \mp U_0 \delta N_{1,2} + N_0 \delta U_{1,2}, \quad (A.3a)$$

$$\delta E_x\big|_{\pm X_0 - 0}^{\pm X_0 + 0} = -\delta N_{1,2} \mp \delta x_{1,2}, \quad (A.3b)$$

$$\delta E_y\big|_{\pm X_0 - 0}^{\pm X_0 + 0} = -ikE_0\delta x_{1,2}. \quad (A.3c)$$

These plasma parameters are found from the linearized continuity equations (1) and the Euler equations (2) for each flow:

$$\delta N_{1,2}(\omega \pm kU_0) = kN_0\delta U_{1,2}, \quad (A.4a)$$

$$-i\Gamma_0(\omega \pm kU_0)\delta v_{x1,x2} = -\delta E_x(\pm X_0) \pm$$
$$\pm U_0\ \delta B_z(\pm X_0) - B_0\ \delta U_{1,2}, \quad (A.4b)$$

$$-i\Gamma_0^3(\omega \pm kU_0)\delta U_{1,2} = -\delta E_y(\pm X_0) +$$
$$+ B_0\ \delta v_{x1,x2}, \quad (A.4c)$$

The effective fields inside the plasma layers at the positions $\pm X_0$ are defined as a linear combination of the inner (A.2a), (A.2b), (A.2c) and the outer fields (A.2e), (A.2f), (A.2g), in the same way, as for the equilibrium fields:

$$\delta E_x(\pm X_0) = Z\delta E_x^{(in)} + (1-Z)\delta E_x^{(out)}, \quad (A.5a)$$

$$\delta E_y(\pm X_0) = Z\delta E_y^{(in)} + (1-Z)\delta E_y^{(out)}, \quad (A.5b)$$

$$\delta B_z(\pm X_0) = Z\delta B_z^{(in)} + (1-Z)\delta B_z^{(out)}, \quad (A.5c)$$

where

$$Z \equiv B(\Gamma_0)/B(\tilde\gamma). \quad (A.5d)$$

It is convenient to express all fluctuations for each flow through the parallel velocities $\delta U_{1,2}$ with the help of (A.4a), (A.4c), and the relation between the transverse velocity and the displacement

$$\delta v_{x1,x2} = -i\omega\delta x_{1,2}. \quad (A.6)$$

From (A.4a), (A.4c), and (A.6) it follows

$$\delta N_{1,2} = \frac{kN_0}{\omega \pm kU_0}\delta U_{1,2}, \quad (A.7)$$

$$\delta x_{1,2} = \pm\frac{\Gamma_0^3(\omega \pm kU_0) + \sqrt{2}D_\pm}{ZkE_0 \pm \omega B_0}\delta U_{1,2}, \quad (A.8)$$

To find the constants $A$, $C$, $D_\pm$ and $F_\pm$ one has to solve the linear system

$$D_\pm \mp A \sinh QX_0 - C \cosh QX_0 = \pm \frac{\omega N_0}{\omega \pm kU_0} \delta U_{1,2}, \tag{A.9a}$$

$$F_\pm - \frac{\omega\sqrt{2}D_\pm}{ZkE_0 \pm \omega B_0} \pm \frac{Ak}{\omega}\sinh QX_0 + \frac{Ck}{\omega}\cosh QX_0 = \left(\mp \frac{kN_0}{\omega \pm kU_0} \mp \frac{\Gamma_0^3(\omega \pm kU_0)}{ZkE_0 \pm \omega B_0}\right)\delta U_{1,2}, \tag{A.9b}$$

$$\sqrt{2}\omega D_\pm \left(1 - \frac{kE_0}{ZkE_0 \pm \omega B_0}\right) \mp \frac{AQ}{\omega}\cosh QX_0 - \frac{CQ}{\omega}\sinh QX_0 = kE_0 \frac{\Gamma_0^3(\omega \pm kU_0)}{ZkE_0 \pm \omega B_0}\delta U_{1,2}. \tag{A.9c}$$

The solution for $A$ and $C$ reads

$$A = \frac{\Phi_1 \Upsilon_{s-}\delta U_1 - \Phi_2 \Upsilon_{s+}\delta U_2}{\Upsilon_{s+}\Upsilon_{c-} + \Upsilon_{s-}\Upsilon_{c+}}, \quad C = \frac{\Phi_1 \Upsilon_{c-}\delta U_1 + \Phi_2 \Upsilon_{c+}\delta U_2}{\Upsilon_{s+}\Upsilon_{c-} + \Upsilon_{s-}\Upsilon_{c+}}, \tag{A.10a}$$

where

$$\Phi_{1,2} = \mp \omega kE_0 \frac{\Gamma_0^3(\omega \pm kU_0)}{ZkE_0 \pm \omega B_0} - \frac{\sqrt{2}\omega^3 N_0}{(\omega \pm kU_0)}\left(1 - \frac{kE_0}{ZkE_0 \pm \omega B_0}\right), \tag{A.10b}$$

$$\Upsilon_{s\pm} = Q \sinh QX_0 - \omega^2\sqrt{2}\left(1 - \frac{kE_0}{ZkE_0 \pm \omega B_0}\right)\cosh QX_0, \tag{A.10c}$$

$$\Upsilon_{c\pm} = Q \cosh QX_0 - \omega^2\sqrt{2}\left(1 - \frac{kE_0}{ZkE_0 \pm \omega B_0}\right)\sinh QX_0. \tag{A.10d}$$

The substitution of the fields (A.5) to (A.4b), gives two equations to define $\delta U_{1,2}$. The full expression is too cumbersome to be written explicitly, however, is may be simplified using the condition $\tilde\gamma_0 \gg 1$, which gives

$$\delta U_{1,2}\left[B_0 - (1-Z)N_0 + \mathcal{P}_\mp \left(\frac{kE_0 \Gamma_0^3(k\pm\omega)^2}{ZkE_0 \pm \omega B_0} - \sqrt{2}\left(1 - \frac{kE_0}{ZkE_0 \pm \omega B_0}\right)\omega^2 N_0\right)\right] -$$

$$-\delta U_{2,1}\mathcal{M}_\pm\left[\frac{kE_0\Gamma_0^3(\omega^2 - k^2)}{ZkE_0 \mp \omega B_0} + \sqrt{2}\omega^2 N_0 \frac{k\pm\omega}{k\mp\omega}\left(1 - \frac{kE_0}{ZkE_0 \mp \omega B_0}\right)\right] = 0, \tag{A.11}$$

where

$$\mathcal{P}_\pm = \left(\frac{\Upsilon_{s\pm}}{\cosh QX_0} + \frac{\Upsilon_{c\pm}}{\sinh QX_0}\right)\bigg/\frac{\Upsilon_{s+}\Upsilon_{c-} + \Upsilon_{s-}\Upsilon_{c+}}{\cosh QX_0 \sinh QX_0}, \tag{A.12a}$$

$$\mathcal{M}_\pm = \left(\frac{\Upsilon_{s\pm}}{\cosh QX_0} - \frac{\Upsilon_{c\pm}}{\sinh QX_0}\right)\bigg/\frac{\Upsilon_{s+}\Upsilon_{c-} + \Upsilon_{s-}\Upsilon_{c+}}{\cosh QX_0 \sinh QX_0}. \tag{A.12b}$$

The model of two thin currents sheets applies if $\tilde\gamma_0 \gtrsim 1$. This condition allows to analyze Eq. (A.11), using the scaling

$$\Gamma_0 \sim \tilde\gamma_0^{3/4}, \tag{A.13a}$$

$$B_0 \sim E_0 \sim N_0 \sim X_0 \sim \tilde\gamma_0^{3/2}, \tag{A.13b}$$

$$1 - U_0 \sim \tilde\gamma_0^{-3/2}. \tag{A.13c}$$

The leading terms in (A.11) are $B_0$ and $(1-Z)N_0$. The solution then may be possible only, if the functions (A.12) compensate the small factors in front of them. For this reason, the expression $\Upsilon_{s+}\Upsilon_{c-} + \Upsilon_{s-}\Upsilon_{c+}$ should be small. This means, that in the limit $\tilde\gamma_0 \to \infty$ the dispersion equation is simplified to (31), and for the more detailed analysis for $\tilde\gamma_0 \gtrsim 1$ Eq. (A.11) should be used.